\documentclass{appolb}
\usepackage{epsfig}

\begin{document}
\title{Can we obtain a ``new femtoscopy''
on the basis of electromagnetic effects?\thanks{Presented at the 
XI Workshop on Particle Correlations and Femtoscopy,
Warsaw Univ. of Technology, Warsaw, Poland, 3-7 November 2015.
}%
}
\author{Andrzej Rybicki$^1$, Antoni Szczurek$^{1,2}$, Miros\l{}aw Kie\l{}bowicz$^1$, Nikolaos Davis$^1$, and Vitalii Ozvenchuk$^1$
\address{$^1$H.~Niewodnicza\'{n}ski Institute of Nuclear Physics, Polish 
 Academy of Sciences, Radzikowskiego 152, 31-342~Krak\'ow, Poland\\
 $^2$University of Rzesz\'ow, Rejtana 16, 35-959 Rzesz\'ow, Poland}
}
\maketitle
\begin{abstract}
We review our studies of spectator-induced electromagnetic (EM) effects on charged pion emission in ultrarelativistic heavy ion collisions. These effects are found to consist in the electromagnetic {charge splitting} of pion directed flow as well as very large distortions in spectra and ratios of produced charged particles. As it emerges from our analysis, they offer sensitivity to the actual distance $d_E$ between the pion formation zone at freeze-out and the spectator matter. As a result, this gives a new possibility of studying the space-time evolution of dense and hot matter created in the course of the collision. Having established that $d_E$ traces the longitudinal evolution of the system and therefore rapidly decreases as a function of pion rapidity, we investigate the latter finding in view of pion feed-over from intermediate resonance production. As a result we obtain a first estimate of the pion decoupling time from EM effects which we compare to existing HBT data. We conclude that spectator-induced EM interactions can serve as a new tool for studying the space-time characteristics and longitudinal evolution of the system. 
 \end{abstract}
\PACS{25.75. -q, 12.38. Mh}

\section{Introduction}
 This paper is largely focused on the known fact that the presence of charged spectator systems in non-central heavy ion collisions gives birth to a sizeable electromagnetic field. In 
our earlier works devoted to different observables like charge-dependent pion yields or directed flow~\cite{twospec,Rybicki-v1,Rybicki-auau} we found that this phenomenon provided new information on the space-time evolution of the system of hot and dense matter created in the reaction. An evident question emerged whether such information could be exploited as a ``new femtoscopy'', based on effects other than particle interferometry.

Below this issue will be reviewed and interpreted in the context of another well known phenomenon which is ``non-direct'' pion production from intermediate hadronic resonances. The very well defined, precise character of the electromagnetic (EM) interaction, in contrast to the non-perturbative strong force, gives the hope of obtaining new space-time information that would be no- or little model-dependent. A first discussion of this item will be made below in the context of the present phenomenological and experimental situation.

\begin{figure}
\begin{center}
\hspace*{-0.2cm}
\includegraphics[height=13.1cm,width=11cm]{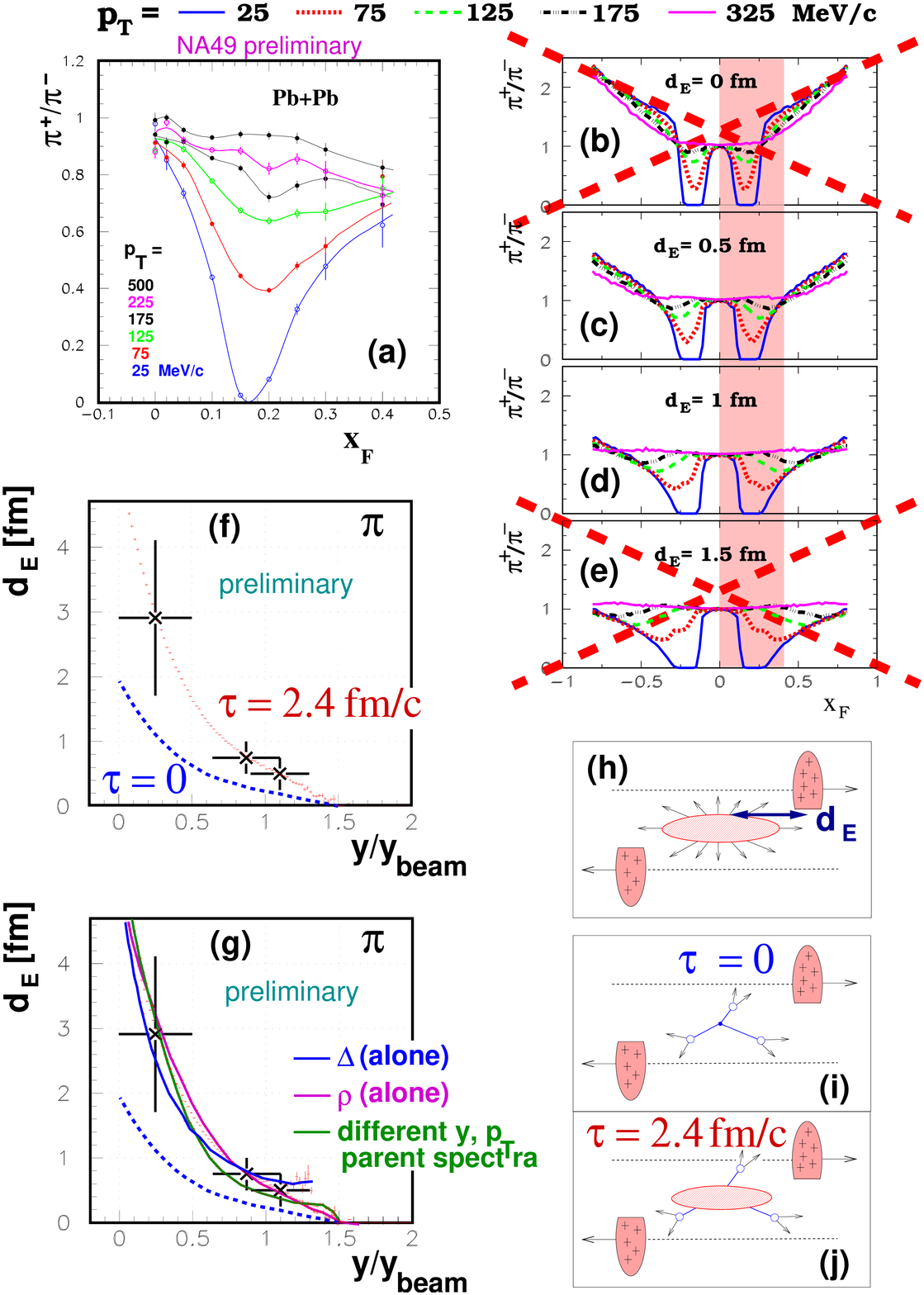}
\end{center}
  \caption{(a) Electromagnetic distortion of $\pi^+/\pi^-$
ratios in peripheral Pb+Pb 
reactions as a function of $x_F$ and $p_T$~\cite{epshep}, compared to (b), (c), (d), (e) corresponding simulated $\pi^+/\pi^-$ ratios for different assumed values of $d_E$~\cite{twospec}. Values of $d_E$ excluded by experimental data 
are indicated in the plot. (f) Dependence of $d_E$ on 
pion rapidity (data points), compared to the results of our Monte Carlo simulation described in section~\ref{characteristics}. 
(g) Same as above, with the addition of results of three supplementary analyses using different input characteristics for the assumed mix of parent resonances.
(h) Schematic definition of $d_E$. (i) Illustration of resonances produced instantly at the moment of the collision ($\tau=0$) and (j) produced from an intermediate system with a given proper lifetime $\tau$.}
 \label{figx}
\end{figure}

\section{Electromagnetic effects}

Three basic datasets on charge dependence of particle production served as input for the present study. These were charged pion ratio NA49~\cite{epshep} and WA98 positive pion directed flow~\cite{wa98} data from Pb+Pb reactions at $\sqrt{s_{NN}}=17.3$~GeV, as well as charge-dependent pion directed flow in Au+Au reactions at $\sqrt{s_{NN}}=7.7$~GeV~\cite{star2014} from the STAR experiment
 (RHIC Beam Energy Scan). 
The data were interpreted by means of a simplified one-point pion emission model, where the propagation of charged pions through the spectator EM field was numerically computed using classical relativistic equations of motion~\cite{twospec}. The results of this analysis appeared sensitive to the ``pion emission distance'' $d_E$ between the pion emission point and the spectator system, see Fig.~\ref{figx}(h). Consequently, the first information on the evolution of $d_E$ as a function of pion rapidity was obtained. An example of our studies for the case of NA49 data is presented in Fig.~\ref{figx}, where the $\pi^+/\pi^-$ ratios measured at high pion rapidities ($x_F>0.1$, panel (a)) were used to eliminate wrong space-time scenarios defined by too small or too large values of $d_E$ (panels (b)-(e)). A summary of the presently available information on $d_E$ as a function of pion rapidity is given in Fig.~\ref{figx}(f) where the lower, intermediate and high values of $y/y_{beam}$ correspond to STAR $v_1$ data, NA49 charged pion ratios, and WA98 $v_1$ data, respectively. The sizeable error bars apparent for the data points in the figure
correspond to our present best estimate of conceptual as well as experimental uncertainties. Improvement in both domains is realistically possible, also in view of the new experimental program proposed by the NA61 experiment at CERN~\cite{spscAddendumPbPb}.

Independently of the above, two distinct features in the rapidity dependence of the distance between the pion at freeze-out and the corresponing position of spectator system are apparent: (a) its decrease as a function of rapidity and (b) the non-zero value of $d_E$ for pions moving at spectator rapidities. Both issues will be addressed below.

\section{Characteristics of space-time evolution of pion production}
\label{characteristics}

A basically model-independent feature of low-momentum transfer processes is ``resonance dominance'' in particle production~\cite{f}. Based on that we performed a simple Monte Carlo study, assuming the bulk of pions to be produced from low-lying resonances like $\Delta(1232)$ or $\rho(770)$ (or other baryonic and mesonic states with roughly similar characteristics). The resulting prediction for $d_E$ as a function of {pion} rapidity
 is shown in Fig.~\ref{figx}(f) as dashed blue curve, for a mix of $\Delta$ and $\rho$ resonances produced {instantly} at the interaction point as illustrated by Fig.~\ref{figx}(i). While this prediction gives a qualitative agreement with the decrease of $d_E$ with $y/y_{beam}$ which we obtained from electromagnetic effects, a quantitative disagreement with these values is evident in spite of large error bars 
associated to
the data points. 

A full quantitative agreement is obtained if, unlike in the precedent case, the parent resonances are not produced instantly, but the existence of an {\em intermediate system prior to resonance formation} is postulated in our study
 as illustrated by Fig.~\ref{figx}(j). This is shown by red crosses in Fig.~\ref{figx}(f). 
Assuming, within sizeable uncertainities, the proper lifetime of this system to be of the order of $\tau\approx 2.4$~fm/c for the considered range of collision energies,
our toy Monte Carlo properly describes not only the decrease of $d_E$ with $y/y_{beam}$ but also its non-zero value at beam rapidity, which then appears as a direct consequence of pion feed-over from resonance production.

{\bf We conclude} that spectator-induced EM effects indeed provide an independent evidence for the existence of an intermediate system of (presumably hot and dense) matter in heavy ion collisions, and can be used to evaluate both its life-time and its longitudinal evolution up to high rapidities. This ``new femtoscopy'' remains completely autonomous from HBT information; the proper lifetime of this ``intermediate system prior to resonance formation'' postulated above corresponds to a final time of pion emission of $\mathbf{5.3\pm 2.2}$~{\bf fm/c} (both value and error are our first and preliminary estimates). For comparison, the compilation of pion decoupling times as presented in the published HBT analysis by the ALICE Collaboration~\cite{alicehbt} gives, for the same range of collision energies, a value of approximatively $\tau_f\approx {6}$~{fm/c}.

Several sources can be claimed to induce uncertainties to the preliminary analysis presented above. A possibly important phenomenological error source is
the lack of full experimental knowledge on all the baryonic/mesonic resonances building up final state pion production. An attempt to estimate the corresponding uncertainty is presented in Fig.~\ref{figx}(g). While our study shown in Fig.~\ref{figx}(f) is performed on the basis of experimental knowledge on yields and kinematical distributions of $\Delta(1232)$ and $\rho(770)$ resonances in p+p collisions~\cite{Rybicki2002dp}, including an estimate of baryon stopping from p+p to A+A reactions as well as proper resonance lifetimes, Breit-Wigner distributions, etc., other analyses assuming different baryonic/mesonic contributions to the resonance mix or different parent $y$ and $p_T$ distributions bring similar results in the pion rapidity range presently available to our studies of EM interactions. Thus we provisionaly conclude that the lack of precise information on resonance distributions is unlikely to constitute a serious obstacle for obtaining new information on the space-time evolution of the system via spectator-induced electromagnetic effects.\\

The authors warmly thank S.~Mr\'owczy\'nski, K.~Redlich, M.~R\'o\.za\'nska, 
L.~Le\'sniak and H.~G.~Fischer for inspiring remarks, discussion and help.
This work was supported by the National Science Centre, Poland
(grants no. 2011/03/B/ST2/02634 and 2014/14/E/ST2/00018).



\begin{thebibliography}{99}

 \bibitem{twospec}
  A.~Rybicki, A.~Szczurek, {\it Phys. Rev.} {\bf C75}, 054903 
(2007).

\bibitem{Rybicki-v1} 
  A.~Rybicki and A.~Szczurek,
  {\it Phys. Rev.} {\bf C87}, 054909 (2013).

\bibitem{Rybicki-auau}
  A.~Rybicki, A.~Szczurek and M.~K\l{}usek-Gawenda,
  {\it Acta Phys.\ Polon.} {\bf B46}, 
  no. 3, 
737 (2015).
 %

 \bibitem{epshep}
A.~Rybicki, PoS(EPS-HEP 2009) 031.
 %

\bibitem{wa98}   %
  WA98 Collab., H.~Schlagheck, 
  {\it Nucl. Phys.} {\bf A663}, 725 (2000).

 %
 %
 %
 %
 %
 %
 %
%
 %
 %


\bibitem{star2014} 
  STAR Collab., L.~Adamczyk {\it et al.}, 
  {\it Phys. Rev. Lett.}  {\bf 112}, 162301 (2014).

\bibitem{spscAddendumPbPb}
NA61 Collab.,
  N.~Abgrall {\it et al.} ,
CERN-SPSC-2015-038.


\bibitem{f}
K.~Fia\l{}kowski and W.~Kittel,
{\it Rept.\ Prog.\ Phys.}  {\bf 46}, 1283 (1983).

 \bibitem{alicehbt} 
ALICE Collab., K.~Aamodt {\it et al.}, {\it Phys. Lett.} {\bf B696}, 328 
(2011).
 %
\bibitem{Rybicki2002dp}
  A.~Rybicki,
  CERN-THESIS-2003-005, and references therein.

\end{thebibliography}
\end{document}